\documentclass[aps,PRB,twocolumn,superscriptaddress,preprintnumbers]{revtex4-2}
\usepackage[T1]{fontenc}
\usepackage{times}
\usepackage{graphicx,color}
\usepackage{amsfonts,amsmath,amssymb,amsbsy}
\usepackage[colorlinks=true, linkcolor=blue, citecolor=blue, urlcolor=blue]{hyperref}
\usepackage{float}

\begin{document}

\title{Tunneling magnetoresistance in a junction made of $X$-wave magnets
with $X=p,d,f,g,i$}
\author{Motohiko Ezawa}
\affiliation{Department of Applied Physics, The University of Tokyo, 7-3-1 Hongo, Tokyo
113-8656, Japan}

\begin{abstract}
We investigate the tunneling magnetoresistance (TMR) of a bilayer system
made of $X$-wave magnets with $X=p,d,f,g,i$, where $X=d,g,i$ corresponds to
altermagnets. A universal analytic formula is derived for the TMR ratio. It
is proportional to $\left\vert J\right\vert /\left( N_{X}\Gamma \right) $
for small $\Gamma $, where $N_{X}$ is the number of the nodes of the $X$%
-wave magnet, $J$ is the strength of the $X$-wave magnet, and $\Gamma $ is
the self-energy. It is contrasted with the TMR ratio made of ferromagnets, where
it is proportional to $J^{2}/\Gamma ^{2}$ for small $\Gamma $. Therefore,
the TMR ratio is larger in ferromagnets for $\left\vert J\right\vert >\Gamma $.
However, the $X$-wave magnets are expected to achieve high-speed and
ultra-dense memory owing to the zero net magnetization. 
\end{abstract}

\date{\today }
\maketitle

\section{Introduction}

Magnetic tunneling junction is a most successful spintronic device\cite%
{Julli,Mood}. It consists of the bilayer ferromagnets spaced by an
insulator. The resistance is low (large) if the directions of spins are
identical (opposite). It is called the tunneling magnetoresistance (TMR).
The spin direction of the memory can be readout by using the TMR. On the
other hand, spintronics based on antiferromagnets is expected to achieve
high speed and ultra dense memories because antiferromagnets have no net
magnetization and free from a stray field\cite%
{Jung,Baltz,Han,Ni,Godin,Kimura,ZhangNeel}. The problem of antiferromagnets
is that it is hard to readout the direction of the N\'{e}el vector.
Altermagnets overcome this problem, though they are antiferromagnet, because
the N\'{e}el vector can be readout by measuring anomalous Hall conductivity%
\cite{Feng,Fak,Tsch,Sato,Leiv,Fak,Tsch,Kagawa,Attias,Seki,Sheoran}, where
their band structures break time-reversal symmetry. There are the $d$-wave, $%
g$-wave and $i$-wave altermagnets\cite{SmejX,SmejX2}. On the other hand, the 
$p$-wave magnet\cite{pwave} and the $f$-wave magnet are also possible\cite%
{SmejX} although they are not altermagnets, where their band structures
preserve time-reversal symmetry. Magnetic tunneling junctions are discussed
in the $d$-wave altermanget\cite{SmejX2,FLiu,Chi,Yao,ZYan,ZYang,YSun} and
the $p$-wave magnet\cite{Brek}.

It would be important to investigate the $d$-wave, $g$-wave and $i$-wave
altermagnets and the $p$-wave and $f$-wave magnets in the universal notion
of the higher symmetric $X$-wave magnets\cite{Planar} with $X=p,d,f,g,i$,
because they share universal physics irrespective of the absence or the
presence of time-reversal symmetry. The characteristic feature is that the
band structure has $N_{X}$ nodes with $N_{X}=1,2,3,4,6$, respectively. So
far, there are no analytic results on the TMR for these $X$-wave magnets.

In this paper, we study the TMR for the bilayer junction made of $X$-wave
magnets and derive analytic formulas. The main result is that the
differential conductivity for the parallel configuration is given by%
\begin{equation}
\lim_{\Gamma \rightarrow 0}\frac{G_{\text{P}}}{4e\pi ^{3}}=\frac{2m\pi ^{2}}{%
\hbar ^{2}\Gamma },
\end{equation}%
while that for the antiparallel configuration is given by%
\begin{equation}
\lim_{\Gamma \rightarrow 0}\frac{G_{\text{AP}}}{4e\pi ^{3}}=\sqrt{\frac{m}{%
2\mu }}\frac{N_{X}\pi ^{2}}{\hbar a\left\vert J\right\vert },
\end{equation}%
where $J$ is the strength of the $X$-wave magnet, $\Gamma $ is the
self-energy, $\mu $ is the chemical potential, $a$ is the lattice constant
and $m$ is the mass of electrons. Hence, the TMR ratio is given by%
\begin{equation}
\lim_{\Gamma \rightarrow 0}\text{TMR}_{\text{ratio}}\equiv \lim_{\Gamma
\rightarrow 0}\frac{G_{\text{P}}-G_{\text{AP}}}{G_{\text{AP}}}=\frac{%
2a\left\vert J\right\vert \sqrt{2m\mu }}{N_{X}\hbar \Gamma }.
\end{equation}%
We compare these analytic formulas to numerical results based on the
tight-binding model.

It is to be contrasted with the TMR ratio based on ferromagnets, where it is
given by%
\begin{equation}
\lim_{\Gamma \rightarrow 0}\text{TMR}_{\text{ratio}}=\frac{2J^{2}}{\Gamma
^{2}}.
\end{equation}%
Therefore, the TMR ratio is larger in ferromagnets for $\left\vert J\right\vert
>\Gamma $. However, the $X$-wave magnets are expected to achieve high-speed
and ultra-dense memory owing to the zero net magnetization.

\section{Model}

\subsection{Continuum model}

The $X$-wave magnet is well described by the two-band Hamiltonian given by%
\cite{GI,Planar}%
\begin{equation}
H=\frac{\hbar ^{2}\mathbf{k}^{2}}{2m}-\mu +Jf_{X}\left( \mathbf{k}\right)
\sigma _{z},  \label{2dHamil}
\end{equation}%
where $f_{X}\left( \mathbf{k}\right) $ characterizes the $X$-wave magnet,
and is given by\cite{SmejX,SmejX2,GI,Planar},%
\begin{align}
f_{s}& =1,  \label{EqX} \\
f_{p}& =k_{x}=ak\cos \phi ,  \label{p-f} \\
f_{d}& =2a^{2}k_{x}k_{y}=a^{2}k^{2}\sin 2\phi ,  \label{d-f} \\
f_{f}& =a^{3}k_{x}\left( k_{x}^{2}-3k_{y}^{2}\right) =a^{3}k^{3}\cos 3\phi ,
\label{f-f} \\
f_{g}& =4a^{4}k_{x}k_{y}\left( k_{x}^{2}-k_{y}^{2}\right) =a^{4}k^{4}\sin
4\phi ,  \label{g-f} \\
f_{i}& =2a^{6}k_{x}k_{y}\left( 3k_{x}^{2}-k_{y}^{2}\right) \left(
k_{x}^{2}-3k_{y}^{2}\right) =a^{6}k^{6}\sin 6\phi ,  \label{i-f}
\end{align}%
where $k_{x}=k\cos \phi $, $k_{y}=k\sin \phi $. The $X$-wave magnet has $%
N_{X}$ nodes in the band structure, where $N_{X}=1,2,3,4,6$ for $X=p,d,f,g,i$%
, respectively. It is the effective Hamiltonian describing the lowest two
bands in the vicinity of the $\Gamma $ point\cite{EzawaPNeel}.

The eigenenergy is given by%
\begin{equation}
\varepsilon _{s}(\mathbf{k})=\frac{\hbar ^{2}\mathbf{k}^{2}}{2m}-\mu
+sJf_{X}\left( \mathbf{k}\right)   \label{Es}
\end{equation}%
with the spin index $s=\pm 1$.

We derive analytic formulas based on the continuum model (\ref{2dHamil}).

\begin{figure}[t]
\centerline{\includegraphics[width=0.48\textwidth]{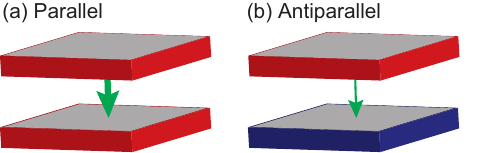}}
\caption{Illustration of a bilayer magnetic tunneling junction made of
magnets. (a) Parallel configuration, where the spin directions are identical
at each lattice site between the two layers. (b) Antiparallel configuration,
where the spin directions are opposite at each lattice site between the two
layers. The green arrow indicates the tunneling current, which is larger in
the parallel configuration.}
\label{FigIllust}
\end{figure}

\subsection{Tight-binding Model}

The tight-binding model\cite{GI,Planar} is constructed from the Hamiltonian (%
\ref{2dHamil}) by replacing $k_{j}\mapsto \sin ak_{j}$ and $k_{j}^{2}\mapsto
2\left( 1-\cos ak_{j}\right) $, where $j=x,y$. The Hamiltonian is defined on
the square lattice for the $p$-wave magnet, the $d$-wave magnet and the $g$%
-wave magnet, while it is defined on the triangular lattice for the $f$-wave
magnet and the $i$-wave magnet.

The tight-binding Hamiltonian is given by%
\begin{equation}
H=H_{\text{Kine}}+H_{X}.  \label{T-Hamil}
\end{equation}%
The kinetic term reads%
\begin{equation}
H_{\text{Kine,Sq}}=\frac{\hbar ^{2}}{ma^{2}}\left( 2-\cos ak_{x}-\cos
ak_{y}\right)  \label{T-sq}
\end{equation}%
on the square lattice, while it reads%
\begin{equation}
H_{\text{Kine,Tri}}=\frac{\hbar ^{2}}{ma^{2}}\left( 3-\cos ak_{x}-\sum_{\pm
}\cos a\frac{k_{x}\pm \sqrt{3}k_{y}}{2}\right)  \label{T-tri}
\end{equation}%
on the triangular lattice. We use Eq. (\ref{T-sq}) for the $p$-wave magnet,
the $d$-wave altermagnet and the $g$-wave altermagnet, while we use Eq. (\ref%
{T-tri}) for the $f$-wave magnet and the $i$-wave altermagnet.

The $X$-wave term $H_{X}$ is given by%
\begin{equation}
H_{p}=J\sigma _{z}\sin ak_{x}
\end{equation}%
for the $p$-wave magnet\cite%
{pwave,Okumura,EzawaPwave,EzawaPNeel,He,Edel,Elliptic}, 
\begin{equation}
H_{d}=J\sigma _{z}\sin ak_{x}\sin ak
\end{equation}%
for the $d$-wave magnet\cite%
{SmejRev,SmejX,SmejX2,Zu2023,Gho,Li2023,EzawaAlter,EzawaMetricC,EzawaVolta},%
\begin{equation}
H_{f}=4J\sigma _{z}\sin ak_{x}\sin \frac{ak_{x}+\sqrt{3}ak_{y}}{2}\sin \frac{%
-ak_{x}+\sqrt{3}ak_{y}}{2}
\end{equation}%
for the $f$-wave magnet\cite{GI,Planar}, 
\begin{equation}
H_{g}=2J\sigma _{z}\sin ak_{x}\sin ak_{y}\left( \cos ak_{y}-\cos
ak_{z}\right) ,
\end{equation}%
for the $g$-wave magnet\cite{GI,Planar}, and%
\begin{align}
H_{i}=& -\frac{16}{3\sqrt{3}}J\sigma _{z}  \notag \\
& \times \sin ak_{x}\sin \frac{ak_{x}+\sqrt{3}ak_{y}}{2}\sin \frac{-ak_{x}+%
\sqrt{3}ak_{y}}{2}  \notag \\
& \times \sin \sqrt{3}ak_{y}\sin \frac{3ak_{x}+\sqrt{3}ak_{y}}{2}\sin \frac{%
-3ak_{x}+\sqrt{3}ak_{y}}{2}
\end{align}%
for the $i$-wave magnet\cite{GI}.

We use the tight-binding model for numerical calculations.

\section{Tunneling magnetoresistance}

We consider a bilayer system made of $X$-wave magnets spaced by an
insulator, as illustrated in Fig.\ref{FigIllust}. The $X$-wave magnet in
each layer is described by the Hamiltonian (\ref{2dHamil}) but may have a
different parameter $J$. We denote it as $J^{\text{T}}$\ in the top layer
and $J^{\text{B}}$\ in the bottom layer. There are two configurations. One
is the parallel configuration, where the spin directions are identical at
each lattice site between the two layers as in Fig.\ref{FigIllust}(a).
Namely, both $X$-wave magnets are described by the Hamiltonian (\ref{2dHamil}%
) together with $J^{\text{T}}=J^{\text{B}}=J$ as in Fig.\ref{FigIllust}(a).
The other is the antiparallel configuration, where the spin directions are
opposite at each lattice site between the two layers as in Fig.\ref%
{FigIllust}(b). Namely, the top $X$-wave magnet is described by the
Hamiltonian (\ref{2dHamil}) while the bottom $X$-wave magnet is described by
the Hamiltonian (\ref{2dHamil}) with the replacement of $J$\ by $-J$, or $J^{%
\text{T}}=-J^{\text{B}}=J$. Large (small) current flows perpendicularly if
the spin directions are identical (opposite) between the two layers. We
calculate the TMR analytically based on the continuum model and numerically
based on the tight-binding model.

The differential conductance $G=dI/dV$ is calculated based on the Green
function\cite{Brek},%
\begin{equation}
\frac{G}{4e\pi ^{3}}=\sum_{s=\pm 1}\sum_{\mathbf{k}_{1},\mathbf{k}%
_{2}}\left\vert T_{\mathbf{k}_{1},\mathbf{k}_{2}}\right\vert ^{2}\text{Tr}%
\left[ \text{Im}\mathcal{G}_{s}^{\text{T}}\left( 0;\mathbf{k}_{1}\right) 
\text{Im}\mathcal{G}_{s}^{\text{B}}\left( 0;\mathbf{k}_{2}\right) \right] ,
\label{G}
\end{equation}%
where $\mathcal{G}_{s}^{\text{T}}$ ($\mathcal{G}_{s}^{\text{B}}$) is the
retarded Green function of the top (bottom) layer defined by%
\begin{equation}
\mathcal{G}_{s}^{\text{T}}\left( \omega ;\mathbf{k}\right) \equiv \mathcal{G}%
_{s}^{\text{B}}\left( \omega ;\mathbf{k}\right) \equiv \mathcal{G}_{s}\left(
\omega ;\mathbf{k}\right) ,
\end{equation}%
for the parallel configuration, and%
\begin{equation}
\mathcal{G}_{s}^{\text{T}}\left( \omega ;\mathbf{k}\right) \equiv \mathcal{G}%
_{s}\left( \omega ;\mathbf{k}\right) ,\quad \mathcal{G}_{s}^{\text{B}}\left(
\omega ;\mathbf{k}\right) \equiv \mathcal{G}_{-s}\left( \omega ;\mathbf{k}%
\right)  \label{GreenAnti}
\end{equation}%
for the antiparallel configuration, where we have defined%
\begin{equation}
\mathcal{G}_{s}\left( \omega ;\mathbf{k}\right) \equiv \frac{1}{\hbar \omega
-\varepsilon _{s}+i\Gamma }
\end{equation}%
with the self-energy $\Gamma $ and Eq.(\ref{Es}) for $\varepsilon _{s}$. The
self-energy is induced for instance by the Coulomb interactions and
impurities. The imaginary part of the Green function is explicitly given by%
\begin{equation}
\text{Im}\mathcal{G}_{s}\left( 0;\mathbf{k}\right) =\frac{\Gamma }{%
\varepsilon _{s}^{2}+\Gamma ^{2}}.
\end{equation}%
It is reduced to the delta function,%
\begin{equation}
\lim_{\Gamma \rightarrow 0}\text{Im}\mathcal{G}_{s}\left( 0;\mathbf{k}%
\right) =-\pi \delta \left( \omega -\varepsilon _{s}\right) ,
\end{equation}%
in the small $\Gamma $ limit.

A comment is in order with respect to the retarded Green functions (\ref%
{GreenAnti}) in the antiparallel configuration. The eigenenergy (\ref{Es})
appears in the Green function, where $J$\ appears only in the combination $Js
$. Hence, the effect of $J^{\text{T}}=-J^{\text{B}}=J$\ is equivalent to
taking $-s$\ instead of $s$\ in the bottom layer, leading to Eq.(\ref%
{GreenAnti}).

\begin{figure*}[t]
\centerline{\includegraphics[width=0.88\textwidth]{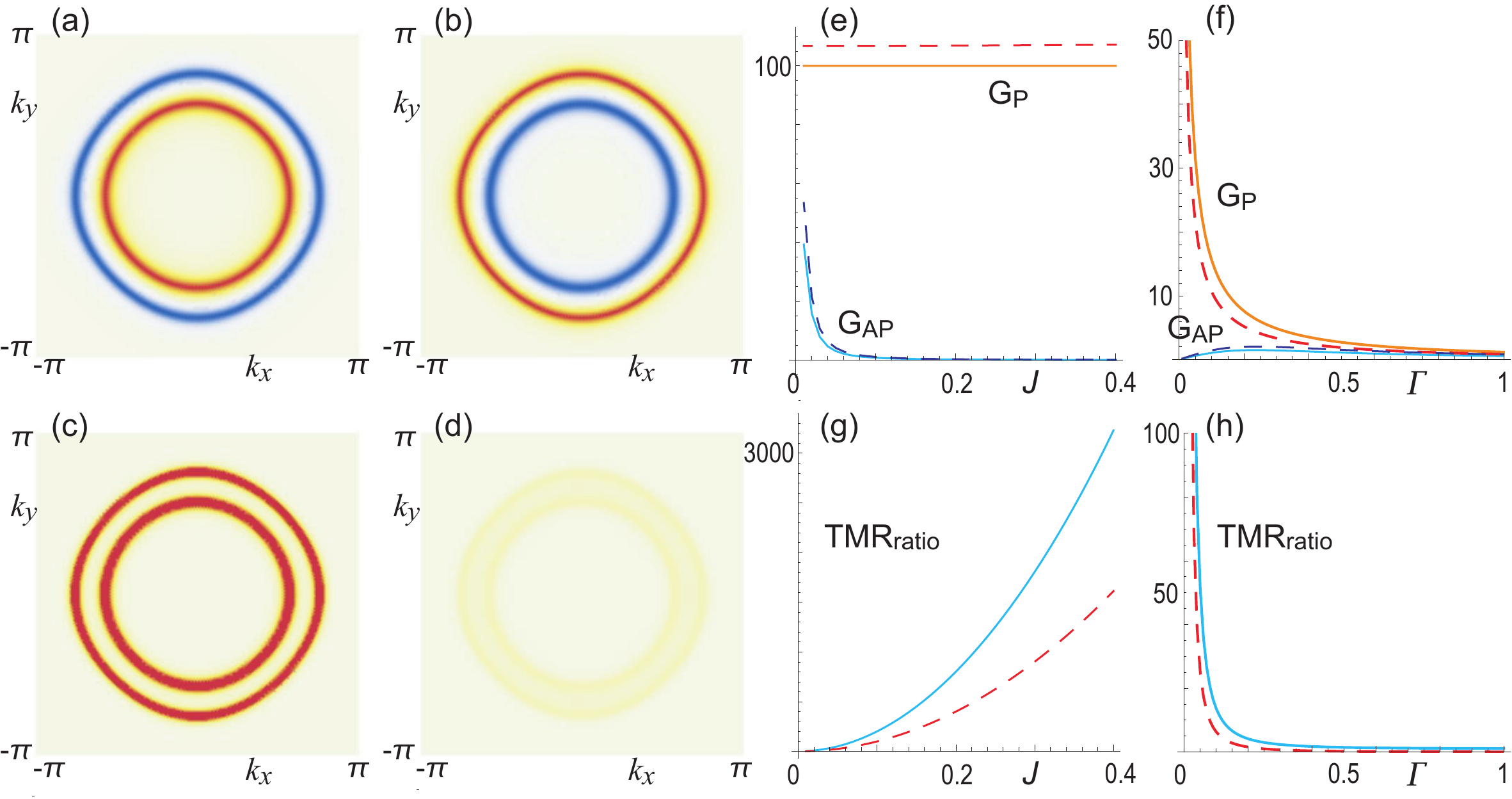}}
\caption{Ferromagnet. (a) Spin density $\left\langle S_{z}\right\rangle $
when $J=0.25\protect\varepsilon _{0}$, and (b) that when $J=-0.25\protect%
\varepsilon _{0}$ in one layer. Red (blue) color indicates up (down) spin.
(c) Overlap $\mathcal{O}_{\text{P}}$ for the parallel configuration, and (d)
overlap $\mathcal{O}_{\text{AP}}$ for the antiparallel configuration. Red
color indicates a large overlap. (e) Differential conductance $G/4e\protect%
\pi ^{5}$ as a function of $J$, and (f) as a function of $\Gamma $. Dashed
red (blue) curves indicate numerical results based on the tight-binding
model for the parallel (antiparallel) configuration. Solid orange (cyan)
curves indicate analytic results based on the continuum model for the
parallel (antiparallel) configuration. (g) TMR ratio as a function of $J$,
and (h) as a function of $\Gamma $. Dashed red (solid cyan) curves indicate
numerical results based on the tight-binding (continuum) model. We have set $%
\hbar ^{2}k_{0}^{2}/2m=\protect\varepsilon _{0},J=0.25\protect\varepsilon %
_{0},$ and $k=1/a$. We have set $\protect\mu =\protect\varepsilon _{0}$ and $%
\Gamma =0.1\protect\varepsilon _{0}$ in (a), (b), (c) and (d), while we have
set $\protect\mu =\protect\varepsilon _{0}/2$ and $\Gamma =0.01\protect%
\varepsilon _{0}$ in (e), (f), (g) and (h).}
\label{FigS}
\end{figure*}

The differential conductance $G_{\text{P}}$ with the parallel configuration
is given by%
\begin{equation}
\frac{G_{\text{P}}}{4e\pi ^{3}}=\sum_{s=\pm 1}\sum_{\mathbf{k}_{1},\mathbf{k}%
_{2}}\left\vert T_{\mathbf{k}_{1},\mathbf{k}_{2}}\right\vert ^{2}\text{Tr}%
\left[ \text{Im}\mathcal{G}_{s}\left( 0;\mathbf{k}_{1}\right) \text{Im}%
\mathcal{G}_{s}\left( 0;\mathbf{k}_{2}\right) \right] ,
\end{equation}%
while the differential conductance $G_{\text{AP}}$ with the antiparallel
configuration is given by%
\begin{equation}
\frac{G_{\text{AP}}}{4e\pi ^{3}}=\sum_{s=\pm 1}\sum_{\mathbf{k}_{1},\mathbf{k%
}_{2}}\left\vert T_{\mathbf{k}_{1},\mathbf{k}_{2}}\right\vert ^{2}\text{Tr}%
\left[ \text{Im}\mathcal{G}_{s}\left( 0;\mathbf{k}_{1}\right) \text{Im}%
\mathcal{G}_{-s}\left( 0;\mathbf{k}_{2}\right) \right] .
\end{equation}%
In Eq.(\ref{G}), $\left\vert T_{\mathbf{k}_{1},\mathbf{k}_{2}}\right\vert
^{2}$ is the tunneling coefficient$\ $given by 
\begin{equation}
\left\vert T_{\mathbf{k}_{1},\mathbf{k}_{2}}\right\vert ^{2}=T_{0}^{2}\delta
\left( \mathbf{k}_{1}-\mathbf{k}_{2}\right)
\end{equation}%
for a bilayer system because the momentum is conserved.

The TMR ratio is defined by using the differential conductances $G_{\text{P}%
} $ ($G_{\text{AP}}$) for the parallel (antiparallel) configuration as\cite%
{Tsy}%
\begin{equation}
\text{TMR}_{\text{ratio}}\equiv \frac{G_{\text{P}}-G_{\text{AP}}}{G_{\text{AP%
}}}.
\end{equation}

In the absence of the magnetic order $J=0$, the differential conductance (%
\ref{G}) is exactly calculated as%
\begin{equation}
\frac{G}{4e\pi ^{3}}=\frac{m\pi }{\hbar ^{2}\Gamma }\left( \frac{2\Gamma \mu 
}{\mu ^{2}+\Gamma ^{2}}+2\arctan \frac{\mu }{\Gamma }+\pi \right) .
\label{Gp0}
\end{equation}%
For $\Gamma /\mu \ll 1$, it is expanded as%
\begin{equation}
\frac{G}{4e\pi ^{3}}=\frac{2m\pi ^{2}}{\hbar ^{2}\Gamma }-\frac{4\pi m\Gamma
^{2}}{3\mu ^{3}}+\frac{8m\pi \Gamma ^{4}}{5\mu ^{5}}-\cdots ,
\end{equation}%
and it is simplified as%
\begin{equation}
\lim_{\Gamma \rightarrow 0}\frac{G}{4e\pi ^{3}}=\frac{2m\pi ^{2}}{\hbar
^{2}\Gamma }
\end{equation}%
in the small $\Gamma $ limit.

We define the overlap between the Fermi surfaces in the top layer and the
bottom layer by%
\begin{eqnarray}
\mathcal{O}_{\text{P}}\left( \mathbf{k}\right)  &\equiv &\sum_{s=\pm 1}\text{%
Tr}\left[ \text{Im}\mathcal{G}_{s}\left( 0;\mathbf{k}\right) \text{Im}%
\mathcal{G}_{s}\left( 0;\mathbf{k}\right) \right] ,  \label{OP} \\
\mathcal{O}_{\text{AP}}\left( \mathbf{k}\right)  &\equiv &\sum_{s=\pm 1}%
\text{Tr}\left[ \text{Im}\mathcal{G}_{s}\left( 0;\mathbf{k}\right) \text{Im}%
\mathcal{G}_{-s}\left( 0;\mathbf{k}\right) \right] .  \label{OAP}
\end{eqnarray}%
With the use of them, the differential conductance (\ref{G}) is simply given
by%
\begin{equation}
\frac{G_{\text{P}}}{4e\pi ^{3}}=\sum_{\mathbf{k}}T_{0}^{2}\mathcal{O}_{\text{%
P}}\left( \mathbf{k}\right) ,\qquad \frac{G_{\text{AP}}}{4e\pi ^{3}}=\sum_{%
\mathbf{k}}T_{0}^{2}\mathcal{O}_{\text{AP}}\left( \mathbf{k}\right) .
\end{equation}

We also define the spin density by%
\begin{equation}
\left\langle S_{z}\left( \mathbf{k}\right) \right\rangle \equiv \text{Im}%
\mathcal{G}_{+}\left( 0;\mathbf{k}\right) -\text{Im}\mathcal{G}_{-}\left( 0;%
\mathbf{k}\right)  \label{SpinD}
\end{equation}%
in each layer.

We calculate these in the following sections for ferromagnet and each $X$%
-wave magnet.

\subsection{Ferromagnet}

For the sake of comparison, we first revisit the ordinary ferromagnetic
magnetic tunneling junction illustrated in Fig.\ref{FigIllust}, and obtain
exact formulas for the TMR. The Hamiltonian is given by setting $f_{X}=1$ in
Eq.(\ref{2dHamil}). The spin density (\ref{SpinD}) is shown in the momentum
space in Fig.\ref{FigS}(a) when $J>0$ and Fig.\ref{FigS}(b) when $J<0$. The
inner Fermi surface is up-spin polarized, while the outer Fermi surface is
down-spin polarized in Fig.\ref{FigS}(a). On the other hand, the inner Fermi
surface is down-spin polarized, while the outer Fermi surface is up-spin
polarized in Fig.\ref{FigS}(b).

\subsubsection{Parallel configuration}

The overlap $\mathcal{O}_{\text{P}}$ for the parallel configuration in Eq.(%
\ref{OP}) is shown in Fig.\ref{FigS}(c). The effect of the ferromagnet is
only to shift the chemical potential from $\mu $ to $\mu -sJ$. By replacing $%
\mu \mapsto \mu -sJ$ in Eq.(\ref{Gp0}), the differential conductance with
the parallel configuration is exactly calculated as%
\begin{eqnarray}
\frac{G_{\text{P}}}{4e\pi ^{3}} &=&\frac{m\pi }{\hbar ^{2}\Gamma }\left( 
\frac{2\Gamma \left( \mu +J\right) }{\left( \mu +J\right) ^{2}+\Gamma ^{2}}%
+2\arctan \frac{\mu +J}{\Gamma }\right.   \notag \\
&&+\left. \frac{2\Gamma \left( \mu -J\right) }{\left( \mu -J\right)
^{2}+\Gamma ^{2}}+2\arctan \frac{\mu -J}{\Gamma }+2\pi \right) .
\end{eqnarray}%
For small $\Gamma $, it is simplified as%
\begin{equation}
\lim_{\Gamma \rightarrow 0}\frac{G_{\text{P}}}{4e\pi ^{3}}=\frac{m\pi ^{2}}{%
2\hbar ^{2}\Gamma }\left( 2+\text{sgn}\left[ \mu -J\right] +\text{sgn}\left[
\mu +J\right] \right) .
\end{equation}%
The differential conductance\textsl{\ }$G_{\text{P}}$ is shown as a function
of $J$ in Fig.\ref{FigS}(e). It is almost flat for small $J$, but it is
slightly different between the tight-binding model and the continuum model
even at $J=0$. This is because the Fermi surface of the tight-binding model
is deformed from the circle.

The differential conductance $G_{\text{P}}$\ is shown as a function of $J$
in Fig.\ref{FigS}(e) and as a function of $\Gamma $ in Fig.\ref{FigS}(f). 

\subsubsection{Antiparallel configuration}

The overlap $\mathcal{O}_{\text{AP}}$ for the antiparallel configuration in
Eq.(\ref{OAP}) is shown in Fig.\ref{FigS}(d). There is almost no overlap for
small $\Gamma $. 

The differential conductance with the antiparallel configuration is exactly
calculated as

\begin{eqnarray}
\frac{G_{\text{AP}}}{4e\pi ^{3}} &=&\frac{\Gamma m}{8\hbar ^{2}J\left(
J^{2}+\Gamma ^{2}\right) }  \notag \\
&&\times \bigg[2J\left( \arctan \frac{J+\mu }{\Gamma }-\arctan \frac{J-\mu }{%
\Gamma }+\pi \right)   \notag \\
&&\hspace{15mm}+\Gamma \log \frac{\Gamma ^{2}+\left( J+\mu \right) ^{2}}{%
\Gamma ^{2}+\left( J-\mu \right) ^{2}}\bigg].
\end{eqnarray}%
It is simplified as%
\begin{equation}
\lim_{\Gamma \rightarrow 0}\frac{G_{\text{AP}}}{4e\pi ^{3}}=\frac{\Gamma m}{%
4\hbar ^{2}J^{2}}\left( 2+\text{sgn}\left[ \mu -J\right] +\text{sgn}\left[
\mu +J\right] \right) 
\end{equation}%
for small $\Gamma $. The differential conductance $G_{\text{AP}}$ is shown
as a function of $J$ in Fig.\ref{FigS}(e) and as a function of $\Gamma $ in
Fig.\ref{FigS}(f). 

\subsubsection{Tunneling magnetoresistance ratio}

The TMR ratio is obtained as%
\begin{equation}
\text{TMR}_{\text{ratio}}=\frac{2J^{2}}{\Gamma ^{2}}-1\simeq \frac{2J^{2}}{%
\Gamma ^{2}}.
\end{equation}%
It is antiproportional to $\Gamma ^{2}$. 

The tunneling magnetoresistance ratio is shown as a function of $J$ in Fig.%
\ref{FigS}(g) and as a function of $\Gamma $ in Fig.\ref{FigS}(h). It
rapidly decreases as the increase of $\Gamma $. It is necessary to make $%
\Gamma $ smaller to achieve larger TMR.

\begin{figure*}[t]
\centerline{\includegraphics[width=0.88\textwidth]{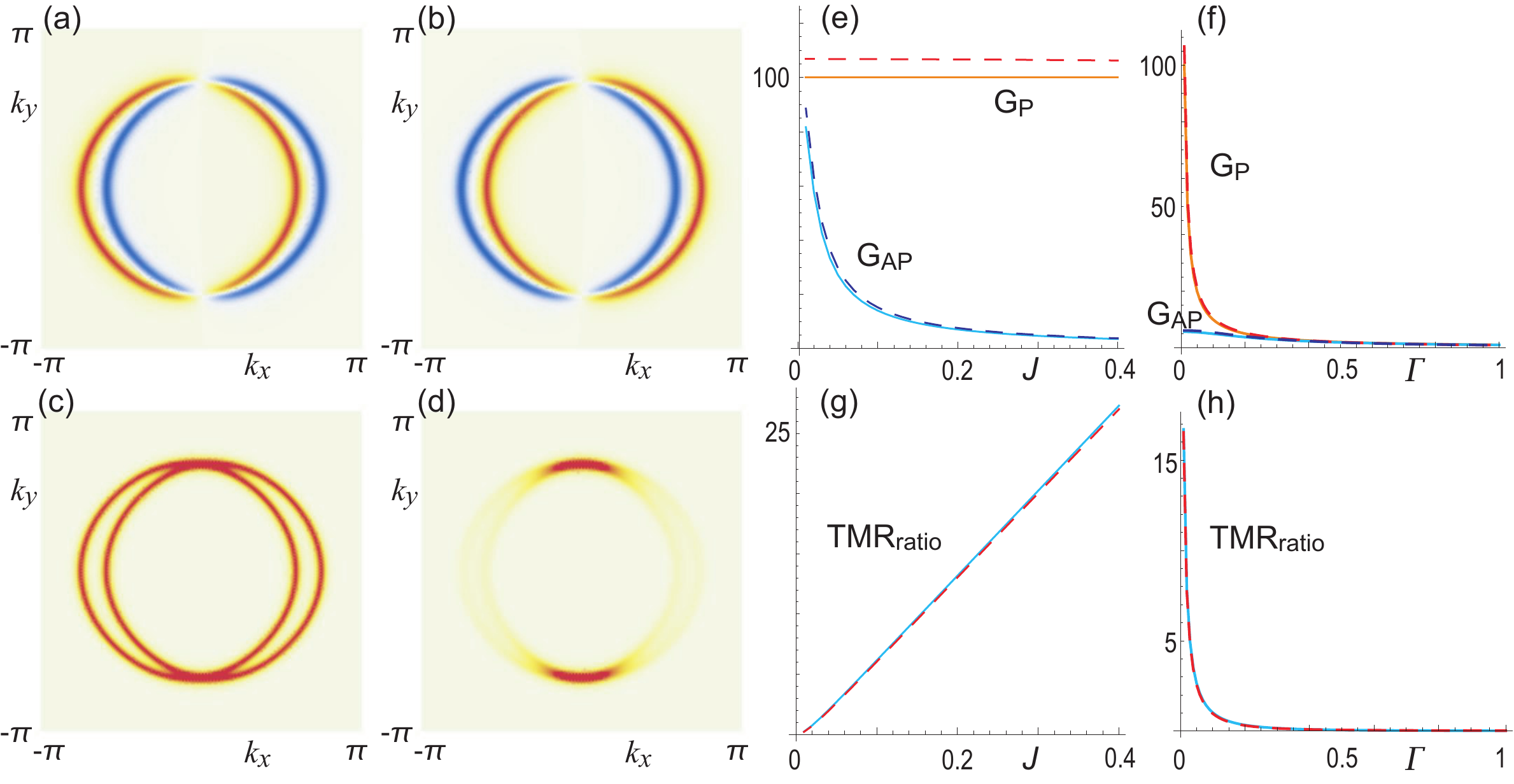}}
\caption{$p$-wave magnet. There are two peaks in (d). See the caption of Fig.%
\protect\ref{FigS}.}
\label{FigP}
\end{figure*}

\subsection{$p$-wave}

The spin density (\ref{SpinD}) is shown in the momentum space in Fig.\ref%
{FigP}(a) when $J>0$ and (b) when $J<0$. The Fermi surfaces shift along the $%
k_{x}$ axis without changing their shapes, where the shifts are opposite
depending on the spins. The overlap $\mathcal{O}_{\text{P}}$ for the
parallel configuration is shown in Fig.\ref{FigP}(c).

The Hamiltonian is rewritten as%
\begin{equation}
\frac{\hbar ^{2}k^{2}}{2m}-\mu \pm Jk_{x}=\frac{\hbar ^{2}k_{x}^{\prime 2}}{%
2m}+\frac{\hbar ^{2}k_{y}^{2}}{2m}-\mu ^{\prime }
\end{equation}%
with%
\begin{equation}
k_{x}^{\prime }\equiv k_{x}\pm mJ/\hbar ^{2},\qquad \mu ^{\prime }\equiv \mu
+\frac{mJ^{2}}{2\hbar ^{2}}.  \label{kmu}
\end{equation}%
The Fermi surface is given by\cite{GI}%
\begin{equation}
\hbar \sqrt{k_{x}^{\prime 2}+k_{y}^{2}}=\sqrt{2m\mu ^{\prime }}.  \label{pFS}
\end{equation}%
It is shifted by $-smJ/\hbar ^{2}$\ for $s=\pm 1$\ in the $p$-wave magnet as
in Fig.\ref{FigP}(a).

\subsubsection{Parallel configuration}

By inserting Eq.(\ref{kmu}) to Eq.(\ref{Gp0}), the differential conductance
is exactly calculated as%
\begin{equation}
\frac{G_{\text{P}}}{4e\pi ^{3}}=\frac{m\pi }{2\hbar ^{2}\Gamma }\left( \frac{%
2\Gamma \mu ^{\prime }}{\mu ^{\prime 2}+\Gamma ^{2}}+2\arctan \frac{\mu
^{\prime }}{\Gamma }+\pi \right) .
\end{equation}%
The differential conductance\textsl{\ }$G_{\text{P}}$ is shown as a function
of $J$ in Fig.\ref{FigP}(e). It depends scarcely on $J$ because the shapes
of the Fermi surfaces do not change as a function of $J$. The differential
conductances $G_{\text{P}}$ is shown as a function of $\Gamma $ in Fig.\ref%
{FigP}(f).

\begin{figure*}[t]
\centerline{\includegraphics[width=0.88\textwidth]{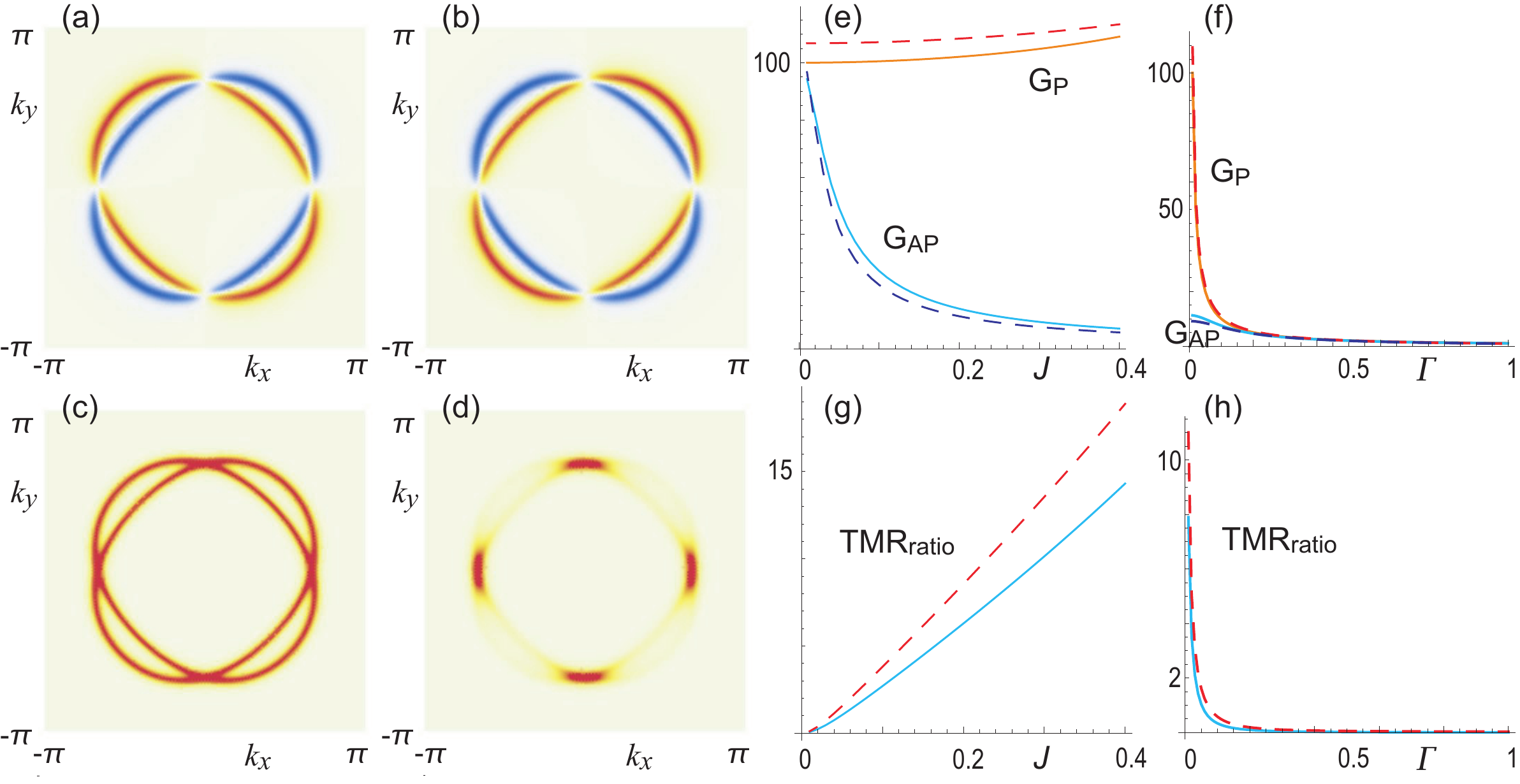}}
\caption{$d$-wave magnet. There are four peaks in (d). See the caption of
Fig.\protect\ref{FigS}.}
\label{FigD}
\end{figure*}

\subsubsection{Antiparallel configuration}

The overlap $\mathcal{O}_{\text{AP}}$ is shown in Fig.\ref{FigP}(d). There
are two peaks, where the two Fermi surfaces with up and down spins overlap.
By solving equations $\varepsilon _{+}=\varepsilon _{-}=0$ with (\ref{Es}),
or%
\begin{equation}
\varepsilon _{\pm }\left( k_{x},k_{y}\right) =\frac{\hbar ^{2}\mathbf{k}^{2}%
}{2m}-\mu \pm Jak_{x}=0,
\end{equation}%
the momenta of the peaks are obtained as%
\begin{equation}
\left( k_{x},k_{y}\right) =\left( 0,\pm \sqrt{2m\mu }\right) .
\end{equation}%
We first take the limit $\Gamma \rightarrow 0$ in the differential
conductance (\ref{G}). It is given by%
\begin{equation}
\frac{G_{\text{AP}}}{4e\pi ^{3}}=\pi ^{2}\sum_{\mathbf{k}}\delta \left(
\varepsilon _{+}\right) \delta \left( \varepsilon _{-}\right) =\pi ^{2}\sum_{%
\mathbf{k}}\frac{\delta \left( k_{x}-k_{+}\right) }{\left\vert \frac{%
\partial \varepsilon _{+}}{\partial k_{x}}\right\vert }\delta \left(
\varepsilon _{-}\right) 
\end{equation}%
with%
\begin{equation}
\frac{\partial \varepsilon _{+}}{\partial k_{x}}=\frac{\hbar ^{2}k_{x}}{m}%
+Ja,
\end{equation}%
and%
\begin{equation}
k_{\pm }=-aJ\frac{m}{\hbar ^{2}}\pm \sqrt{2m\mu +a^{2}J^{2}\frac{m^{2}}{%
\hbar ^{2}}-k_{y}^{2}}.
\end{equation}%
It is further calculated as%
\begin{equation}
\frac{G_{\text{AP}}}{4e\pi ^{3}}=\sum_{k_{y},\pm }\frac{\pi ^{2}\delta
\left( k_{y}\pm \sqrt{2m\mu }\right) }{\left\vert \left. \frac{\partial
\varepsilon _{+}}{\partial k_{x}}\right\vert _{k_{x}=k_{\pm }}\right\vert
\left\vert \left. \frac{\partial \varepsilon _{-}}{\partial k_{y}}%
\right\vert _{k_{x}=k_{\pm }}\right\vert }=\sqrt{\frac{2m}{\mu }}\frac{\pi
^{2}}{\hbar a\left\vert J\right\vert }.  \notag
\end{equation}%
By requiring the Lorentzian form, we recover $\Gamma $ in this equation as%
\begin{equation}
\frac{G_{\text{AP}}}{4e\pi ^{3}}=\sqrt{\frac{2m}{\mu }}\frac{\pi ^{2}}{\hbar 
\sqrt{a^{2}J^{2}+\frac{\hbar ^{2}}{2m\mu }\Gamma ^{2}}},  \label{pGap}
\end{equation}%
where we have determined the coefficient of $\Gamma $ to recover the result
with $J=0$ as%
\begin{equation}
\lim_{J\rightarrow 0}\frac{G_{\text{AP}}}{4e\pi ^{3}}=\frac{2m\pi ^{2}}{%
\hbar ^{2}\Gamma }.
\end{equation}%
The differential conductance $G_{\text{AP}}$ is shown as a function of $J$
in Fig.\ref{FigP}(e). It monotonically decreases as the increase of $J$
because the overlap between the Fermi surfaces with up and down spins
decreases as the increase of $J$. The differential conductances $G_{\text{AP}%
}$ is shown as a function of $\Gamma $ in Fig.\ref{FigP}(f). The numerical
result and the analytic result well agree.

\subsubsection{Tunneling magnetoresistance ratio}

The TMR ratio is obtained as%
\begin{equation}
\text{TMR}_{\text{ratio}}=\frac{\sqrt{2m\mu a^{2}J^{2}+\hbar ^{2}\Gamma ^{2}}%
}{\hbar \Gamma }-1.
\end{equation}%
It is simplified as%
\begin{equation}
\lim_{\Gamma \rightarrow 0}\text{TMR}_{\text{ratio}}=\frac{\sqrt{2m\mu }%
a\left\vert J\right\vert }{\hbar \Gamma }
\end{equation}%
for small $\Gamma $.

\begin{figure*}[t]
\centerline{\includegraphics[width=0.88\textwidth]{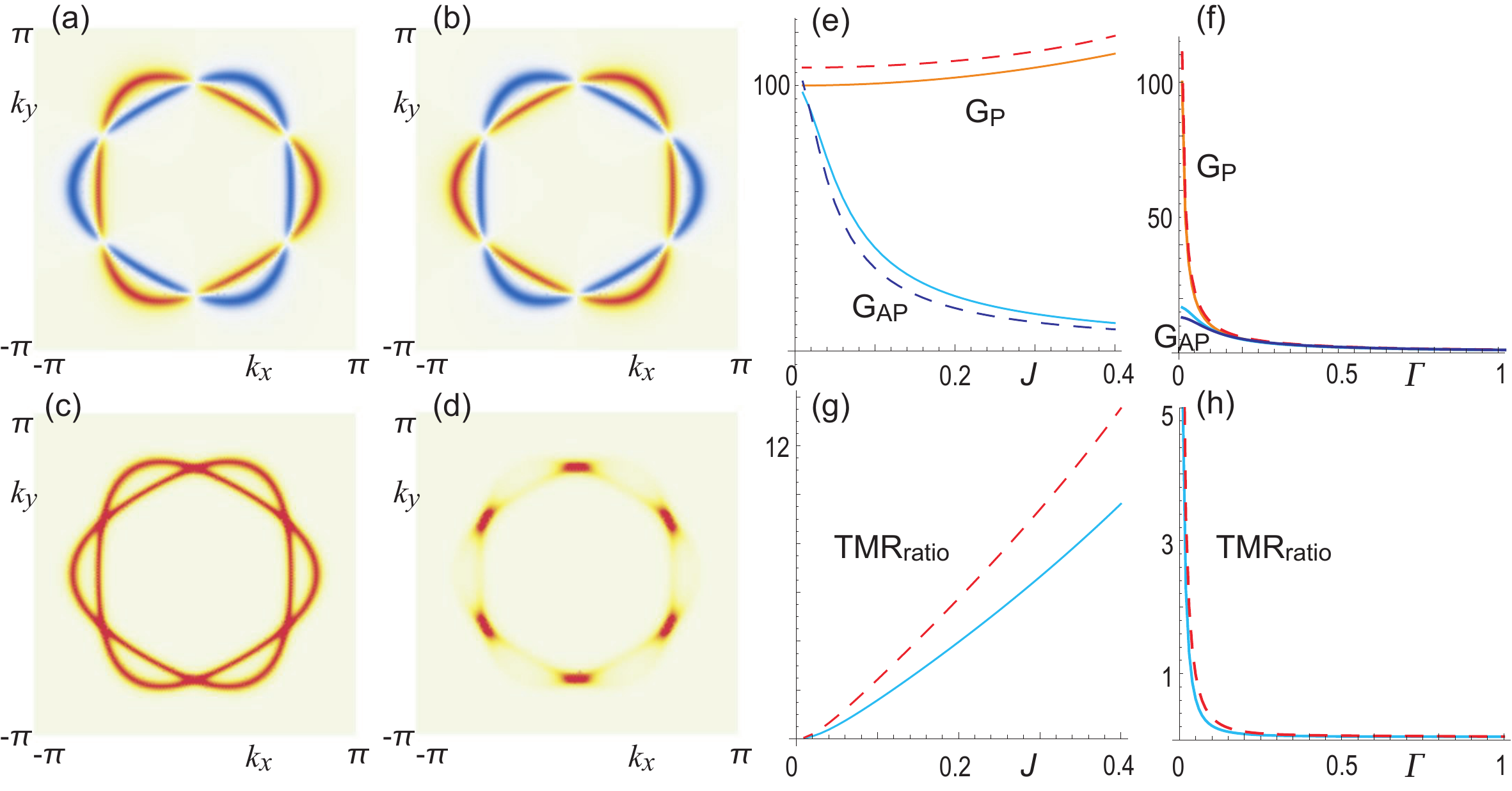}}
\caption{$f$-wave magnet. There are six peaks in (d). See the caption of Fig.%
\protect\ref{FigS}.}
\label{FigF}
\end{figure*}

The TMR ratio is shown as a function of $J$ in Fig.\ref{FigP}(g). It
monotonically increases as a function of $J$. They rapidly decrease as the
increase of $\Gamma $. The TMR ratio is shown as a function of $\Gamma $ in
Fig.\ref{FigP}(h). It also rapidly decrease as the increase of $\Gamma $.

\subsection{$d$-wave}

By solving $\varepsilon _{s}=0$ in Eq.(\ref{Es}), the Fermi surface is
analytically obtained\cite{GI} as a function of $\phi $,%
\begin{equation}
\hbar k_{s}^{\text{F}}\left( \phi \right) =\sqrt{\frac{2\mu }{\frac{1}{m}%
+sa^{2}J\sin 2\phi }}.  \label{dphi}
\end{equation}%
The spin density (\ref{SpinD}) is shown in the momentum space in Fig.\ref%
{FigD}(a) when $J>0$ and (b) when $J<0$. The Fermi surface is deformed to an
elliptic shape for each spin\emph{\ }$s=\pm 1$ in the $d$-wave magnet.

\subsubsection{Parallel configuration}

The overlaps $\mathcal{O}_{\text{P}}$\ for the parallel configuration is
shown in Fig.\ref{FigD}(c). The differential conductance is exactly
calculated as%
\begin{equation}
\frac{G_{\text{P}}}{4e\pi ^{3}}=\frac{m\pi }{2\hbar ^{2}\Gamma }\frac{1}{%
\sqrt{1-4\frac{m^{2}a^{2}J^{2}}{\hbar ^{4}}}}\left( \frac{2\Gamma \mu }{\mu
^{2}+\Gamma ^{2}}+2\arctan \frac{\mu }{\Gamma }+\pi \right) .
\end{equation}%
It is valid only for $4m^{2}J^{2}/\hbar ^{4}<1$, which is identical to the
condition that the parabola is convex upward, or%
\begin{equation}
\frac{\hbar ^{2}\mathbf{k}^{2}}{2m}>\left\vert J\right\vert \mathbf{k}^{2}.
\end{equation}%
It is simplified as%
\begin{equation}
\lim_{\Gamma \rightarrow 0}\frac{G_{\text{P}}}{4e\pi ^{3}}=\frac{2m\pi ^{2}}{%
\hbar ^{2}\Gamma }\frac{1}{\sqrt{1-4\frac{m^{2}a^{2}J^{2}}{\hbar ^{4}}}}
\end{equation}%
for small $\Gamma $. The differential conductances $G_{\text{P}}$ is shown
as a function of $J$ in Fig.\ref{FigD}(e) and as a function of $\Gamma $ in
Fig.\ref{FigD}(f) .

\subsubsection{Antiparallel configuration}

The overlaps $\mathcal{O}_{\text{AP}}$\ for the antiparallel configuration
is shown in Fig.\ref{FigD}(d), respectively. There are four peaks at $\left(
k_{x},k_{y}\right) =\left( 0,\pm \sqrt{2m\mu }\right) ,\left( \pm \sqrt{%
2m\mu },0\right) $, where the two Fermi surfaces with up and down spins are
overlapped. Then, the differential conductance is twice of that of the $p$%
-wave magnet (\ref{pGap}), and given by 
\begin{equation}
\frac{G_{\text{AP}}}{4e\pi ^{3}}=\sqrt{\frac{2m}{\mu }}\frac{2\pi ^{2}}{%
\hbar a\left\vert J\right\vert }.
\end{equation}%
By requiring the Lorentzian form, we recover $\Gamma $ in this equation as%
\begin{equation}
\frac{G_{\text{AP}}}{4e\pi ^{3}}=\sqrt{\frac{2m}{\mu }}\frac{\pi ^{2}}{\hbar 
\sqrt{\frac{a^{2}J^{2}}{4}+\frac{\hbar ^{2}}{2m\mu }\Gamma ^{2}}}.
\end{equation}%
The differential conductances $G_{\text{AP}}$ is shown as a function of $J$
in Fig.\ref{FigD}(e) and as a function of $\Gamma $ in Fig.\ref{FigD}(f) .

\subsubsection{Tunneling magnetoresistance ratio}

The TMR ratio is shown as a function of $J$ in Fig.\ref{FigD}(g). The
tunneling magnetoresistance ratio is shown as a function of $\Gamma $ in Fig.%
\ref{FigD}(h).

\begin{figure*}[t]
\centerline{\includegraphics[width=0.88\textwidth]{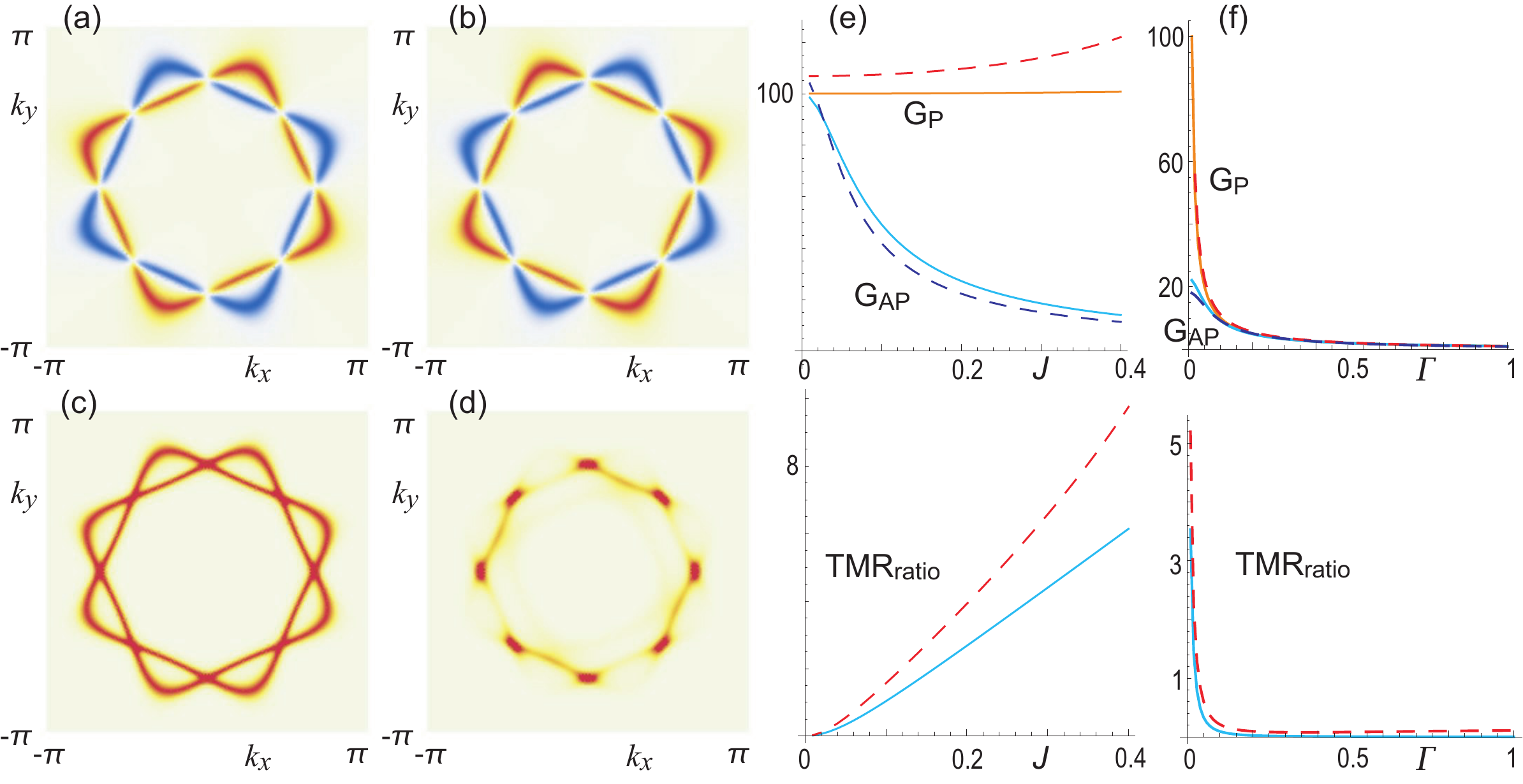}}
\caption{$g$-wave magnet. There are eight peaks in (d). See the caption of
Fig.\protect\ref{FigS}.}
\label{FigG}
\end{figure*}

\subsection{$f$-wave}

By solving $\varepsilon _{s}=0$ in Eq.(\ref{Es}), the Fermi surface is
described by\cite{GI}%
\begin{equation}
\hbar k_{s}^{\text{F}}\left( \phi \right) =\sqrt{2\mu m}-2m^{2}\mu
sa^{3}J\cos 3\phi
\end{equation}%
up to the first order in $J$. The spin density (\ref{SpinD}) is shown in the
momentum space in Fig.\ref{FigF}(a) when $J>0$ and (b) when $J<0$. The Fermi
surface is deformed to a triangular shape for each spin\emph{\ }$s=\pm 1$ in
the $f$-wave magnet.

\subsubsection{Parallel configuration}

The overlap $\mathcal{O}_{\text{P}}$ for the parallel configuration is shown
in Fig.\ref{FigF}(c). The differential conductance is given by%
\begin{equation}
\frac{G_{\text{P}}}{4e\pi ^{3}}=\frac{2m\pi ^{2}}{\hbar ^{2}\Gamma }\left(
1+12\left( \frac{m}{\hbar ^{2}}\right) ^{3}\mu a^{2}J^{2}\right)
\end{equation}%
up to the second order in $J$ for small $\Gamma $. The differential
conductances $G_{\text{P}}$ is shown as a function of $J$ in Fig.\ref{FigF}%
(e) and as a function of $\Gamma $ in Fig.\ref{FigF}(f) .

\subsubsection{Antiparallel configuration}

The overlap $\mathcal{O}_{\text{AP}}$ for the antiparallel configuration is
shown in Fig.\ref{FigF}(d). There are six peaks at $\left(
k_{x},k_{y}\right) =\sqrt{2m\mu }\left( \cos \frac{n\pi }{3},\sin \frac{n\pi 
}{3}\right) $ with $n=0,1,2,3,4,5$, where the two Fermi surfaces with up and
down spins are overlapped. Then, the differential conductance is three times
larger than that of the $p$-wave magnet (\ref{pGap}), and given by 
\begin{equation}
\frac{G_{\text{AP}}}{4e\pi ^{3}}=\sqrt{\frac{2m}{\mu }}\frac{3\pi ^{2}}{%
\hbar a\left\vert J\right\vert }.
\end{equation}%
By assuming the Lorentzian form, we recover $\Gamma $ as%
\begin{equation}
\frac{G_{\text{AP}}}{4e\pi ^{3}}=\sqrt{\frac{2m}{\mu }}\frac{\pi ^{2}}{\hbar 
\sqrt{\frac{a^{2}J^{2}}{9}+\frac{\hbar ^{2}}{2m\mu }\Gamma ^{2}}}.
\end{equation}

The differential conductances $G_{\text{AP}}$ is shown as a function of $J$
in Fig.\ref{FigF}(e) and as a function of $\Gamma $ in Fig.\ref{FigF}(f) .

\subsubsection{Tunneling magnetoresistance ratio}

The tunneling magnetoresistance ratio is shown as a function of $J$ in Fig.%
\ref{FigF}(g). The tunneling magnetoresistance ratio is shown as a function
of $\Gamma $ in Fig.\ref{FigF}(h).

\begin{figure*}[t]
\centerline{\includegraphics[width=0.88\textwidth]{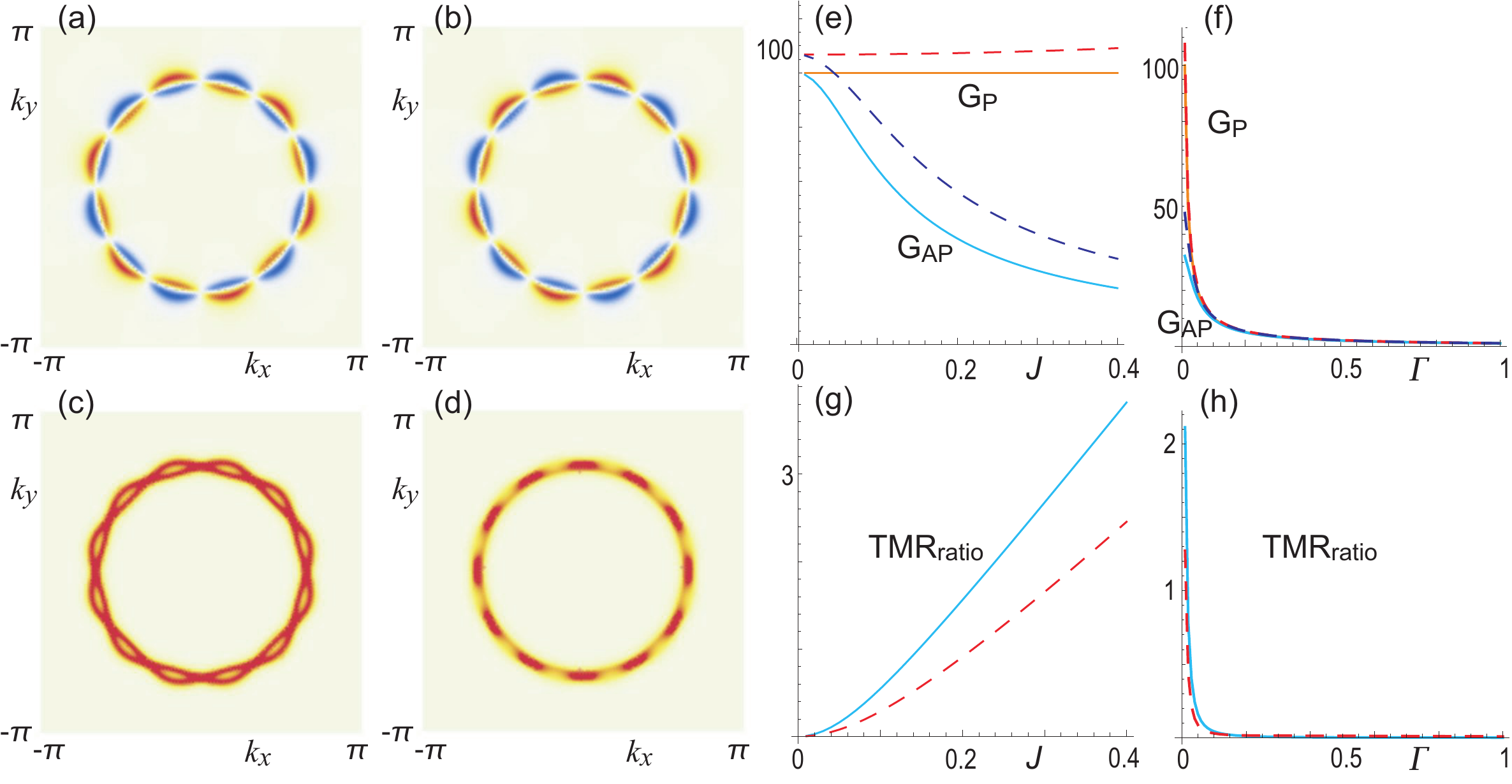}}
\caption{$i$-wave magnet. There are twelve peaks in (d). See the caption of
Fig.\protect\ref{FigS}.}
\label{FigI}
\end{figure*}

\subsection{$g$-wave}

By solving $\varepsilon _{s}=0$ in Eq.(\ref{Es}), the Fermi surface is
described by\cite{GI}%
\begin{equation}
\hbar k_{s}^{\text{F}}\left( \phi \right) =\sqrt{\frac{-1+\sqrt{1+4\mu
sa^{4}Jm^{2}\sin 4\phi }}{sJm\sin 4\phi }}.
\end{equation}%
The spin density (\ref{SpinD}) is shown in the momentum space in Fig.\ref%
{FigG}(a) when $J>0$ and (b) when $J<0$. The Fermi surface is deformed to a
square shape for each spin\emph{\ }$s=\pm 1$ in the $g$-wave magnet.

\subsubsection{Parallel configuration}

The overlap $\mathcal{O}_{\text{P}}$ for the parallel configuration is shown
in Fig.\ref{FigG}(c). The differential conductance is given by 
\begin{equation}
\lim_{\Gamma \rightarrow 0}\frac{G_{\text{P}}}{4e\pi ^{3}}=\frac{2m\pi ^{2}}{%
\hbar ^{2}\Gamma }\left( 1+3\left( \frac{m}{\hbar ^{2}}\right) ^{4}\mu
^{2}J^{2}\right)
\end{equation}%
up to the second order in $J$ for small $\Gamma $. The differential
conductances $G_{\text{P}}$ is shown as a function of $J$ in Fig.\ref{FigG}%
(e) and as a function of $\Gamma $ in Fig.\ref{FigG}(f) .

\subsubsection{Antiparallel configuration}

The overlap $\mathcal{O}_{\text{AP}}$ for the antiparallel configuration is
shown in Fig.\ref{FigG}(d). There are eight peaks at $\left(
k_{x},k_{y}\right) =\sqrt{2m\mu }\left( \cos \frac{n\pi }{4},\sin \frac{n\pi 
}{4}\right) $ with $n=0,1,\cdots ,7$. Then, the differential conductance is
four times larger than that of the $p$-wave magnet (\ref{pGap}), and given
by 
\begin{equation}
\frac{G_{\text{AP}}}{4e\pi ^{3}}=\sqrt{\frac{2m}{\mu }}\frac{4\pi ^{2}}{%
\hbar a\left\vert J\right\vert }.
\end{equation}%
By requiring the Lorentzian form, we recover $\Gamma $ in this equation as%
\begin{equation}
\frac{G_{\text{AP}}}{4e\pi ^{3}}=\sqrt{\frac{2m}{\mu }}\frac{\pi ^{2}}{\hbar 
\sqrt{\frac{a^{2}J^{2}}{16}+\frac{\hbar ^{2}}{2m\mu }\Gamma ^{2}}}.
\end{equation}%
The differential conductances $G_{\text{AP}}$ is shown as a function of $J$
in Fig.\ref{FigG}(e) and as a function of $\Gamma $ in Fig.\ref{FigG}(f) .

\subsubsection{Tunneling magnetoresistance ratio}

The tunneling magnetoresistance ratio is shown as a function of $J$ in Fig.%
\ref{FigG}(g). The tunneling magnetoresistance ratio is shown as a function
of $\Gamma $ in \ref{FigG}(h).

\subsection{$i$-wave}

By solving $\varepsilon _{s}=0$ in Eq.(\ref{Es}), the Fermi surface is
described by\cite{GI}%
\begin{equation}
\hbar k_{s}^{\text{F}}\left( \phi \right) =\sqrt{2m\mu }\left( 1-2\mu
^{2}m^{3}sa^{6}J\sin 6\phi \right)
\end{equation}%
up to the first order in $J$. The spin density (\ref{SpinD}) is shown in the
momentum space in Fig.\ref{FigI}(a) when $J>0$ and (b) when $J<0$. The Fermi
surface is deformed to a hexagonal shape for each spin\emph{\ }$s=\pm 1$ in
the $i$-wave magnet.

\subsubsection{Parallel configuration}

The overlap $\mathcal{O}_{\text{P}}$ for the parallel configuration is shown
in Fig.\ref{FigI}(c). The differential conductance is given by 
\begin{equation}
\lim_{\Gamma \rightarrow 0}\frac{G_{\text{P}}}{4e\pi ^{3}}=\frac{2m\pi ^{2}}{%
\hbar ^{2}\Gamma }
\end{equation}%
up to the first order in $J$ for small $\Gamma $. The differential
conductances $G_{\text{P}}$ is shown as a function of $J$ in Fig.\ref{FigI}%
(e) and as a function of $\Gamma $ in Fig.\ref{FigI}(f) .

\subsubsection{Antiparallel configuration}

The overlap $\mathcal{O}_{\text{AP}}$ for the antiparallel configuration is
shown in Fig.\ref{FigI}(d). There are twelve peaks at $\left(
k_{x},k_{y}\right) =\sqrt{2m\mu }\left( \cos \frac{n\pi }{6},\sin \frac{n\pi 
}{6}\right) $ with $n=0,1,\cdots ,11$. Then, the differential conductance is
6 times larger than that of the $p$-wave magnet (\ref{pGap}), and given by 
\begin{equation}
\frac{G_{\text{AP}}}{4e\pi ^{3}}=\sqrt{\frac{2m}{\mu }}\frac{6\pi ^{2}}{%
\hbar a\left\vert J\right\vert }.
\end{equation}%
By requiring a Lorentzian form, we recover $\Gamma $ in this equation as%
\begin{equation}
\frac{G_{\text{AP}}}{4e\pi ^{3}}=\sqrt{\frac{2m}{\mu }}\frac{\pi ^{2}}{\hbar 
\sqrt{\frac{a^{2}J^{2}}{36}+\frac{\hbar ^{2}}{2m\mu }\Gamma ^{2}}}.
\end{equation}%
The differential conductances $G_{\text{AP}}$ is shown as a function of $J$
in Fig.\ref{FigI}(e) and as a function of $\Gamma $ in Fig.\ref{FigI}(f) .

\subsubsection{Tunneling magnetoresistance ratio}

The tunneling magnetoresistance ratio is shown as a function of $J$ in Fig.%
\ref{FigI}(g). The tunneling magnetoresistance ratio is shown as a function
of $\Gamma $ in \ref{FigI}(h).

\section{Materials}

It was theoretically proposed\cite{pwave} that CeNiAsO is a $p$-wave magnet
and experimentally observed\cite{HZhou}. It was also theoretically proposed%
\cite{Roy} that a $p$-wave magnet is realized in graphene by introducing the
spin nematic order. Experiments on $p$-wave magnets\ were reported\cite%
{Yamada} in Gd$_{3}$Ru$_{4}$Al$_{12}$, and reported\cite{Comin} in NiI$_{2}$%
. A $d$-wave magnet is realized in RuO$_{2}$\cite{Ahn,SmeRuO,Tsch,Fed,Lin},
Mn$_{5}$Si$_{3}$\cite{Leiv,Reich} FeSb$_{2}$\cite{Mazin}, organic materials%
\cite{Naka}, perovskite materials\cite{NakaB,NakaRev}, and twisted magnetic
Van der Waals bilayers\cite{YLiu}. It was theoretically proposed\cite{Roy}
that an $f$-wave magnet is realized in the rhombohedral trilayer graphene by
introducing the spin nematic order. A $g$-wave altermagnet and an $i$-wave
altermagnet are realized in twisted magnetic Van der Waals bilayers\cite%
{YLiu}.

\section{Discussion}

We have analytically derived a formula of the TMR ratio based on the $X$%
-wave magnet, which is proportional to $\left\vert J\right\vert /\left(
N_{X}\Gamma \right) $. On the other hand, the TMR ratio based on
ferromagnets is proportional to $J^{2}/\Gamma ^{2}$. Hence, the TMR ratio is
larger in ferromagnets for $\left\vert J\right\vert >\Gamma $. However, the $%
X$-wave magnets are expected to achieve high-speed and ultra-dense memory
owing to the zero net magnetization.

The author is grateful to M. Hirschberger and A. Tsukazaki for helpful
discussions on the subject. This work is supported by CREST, JST (Grants No.
JPMJCR20T2) and Grants-in-Aid for Scientific Research from MEXT KAKENHI
(Grant No. 23H00171).

\end{document}